\documentstyle[12pt,fleqn,colordvi]{article}
\input epsf
\textBlack
\newcommand{\Frac}[2]{\frac{{\displaystyle #1}}{{\displaystyle #2}}}

\def\ga{\mathrel{\mathpalette\fun >}}
\def\fun#1#2{\lower3.6pt\vbox{\baselineskip0pt\lineskip.9pt
        \ialign{$\mathsurround=0pt#1\hfill##\hfil$\crcr#2\crcr\sim\crcr}}}
\begin{document}

\setcounter{page}{0}
\thispagestyle{empty}

\begin{center}
{\LARGE \bf Exotic massive hadrons and \\  \vspace{6pt}
ultra-high energy cosmic rays} 
\baselineskip=16pt
\vspace{0.75cm}

{\bf Ivone F.\ M.\  Albuquerque}\footnote{Electronic mail: 
		{\tt ifreire@mafalda.uchicago.edu}}\\
{\em Department of Astronomy and Astrophysics, Enrico Fermi Institute\\ 
The University of Chicago, Chicago, Illinois~~60637}\\
\vspace{0.4cm}
{\bf Glennys R.\ Farrar}\footnote{Electronic mail: 
		{\tt farrar@farrar.rutgers.edu}}\\
{\em Department of Physics and Astronomy\\
Rutgers University, Piscataway, New Jersey~~08855}\\
\vspace{0.4cm}
{\bf Edward W.\ Kolb}\footnote{Electronic mail: 
	{\tt rocky@rigoletto.fnal.gov}}\\
{\em NASA/Fermilab Astrophysics Center\\
Fermi National Accelerator Laboratory, Batavia, Illinois~~60510, and\\
Department of Astronomy and Astrophysics, Enrico Fermi Institute\\
The University of Chicago, Chicago, Illinois~~60637}\\
\end{center}
\baselineskip=24pt

\begin{quote}
\hspace*{1em} We investigate the proposal that primary cosmic rays of
energy above the Greisen--Zatsepin--Kuzmin cutoff are exotic massive
strongly interacting particles (uhecrons).  We study the properties of
air showers produced by uhecrons and find that masses in excess of
about 50 GeV are inconsistent with the highest energy event observed.
We also estimate that with sufficient statistics a uhecron of mass
as low as 10 GeV may be distinguished from a proton.
\vspace*{8pt}

PACS number(s): 96.40, 98.70, 11.30.P

\end{quote}

\renewcommand{\thefootnote}{\arabic{footnote}}
\addtocounter{footnote}{-3}

\newpage

\setcounter{page}{1}

\baselineskip=24pt

\section{Introduction}
\hspace*{1em} About thirty years ago it was pointed out that
sufficiently energetic cosmic rays will lose energy scattering with
the cosmic microwave background (CMB) or radio wavelength
background radiation (RBR).  Greisen \cite{Greisen} and Zatsepin and Kuzmin
\cite{ZK} (GZK) noticed that above threshold for pion production a
cosmic-ray proton will lose energy by pion photoproduction.
Sufficiently energetic nuclei also have a limited range before
suffering photodissociation.  The mean-free-path for a proton with
energy above pion production threshold is less than a dozen Mpc, and
the mean--free--path for a nucleus above threshold for
photodissociation is even shorter\footnote{However, the original
estimates of the infrared and far infrared background have been
recently reexamined, leading to a possibility of a larger
photodissociation mean free path \cite{stecker}.}.  

With the assumption that very high energy cosmic rays originate from
sources more distant than this, Greisen and Zatsepin and Kuzmin predicted
that there should be a maximum observed energy in the cosmic-ray spectrum.
This maximum energy, known as the GZK cutoff, is usually taken to be
around $5 \times 10^{19}$eV.  Surprisingly, today there are more than
20 events observed with energy in excess of the GZK limit, and 8
events\footnote{This number does not include events from southern
hemisphere detectors.} above $10^{20}$eV \cite{vr,hp,yak,flys,akeno}.
The slope of the energy spectrum appears to change around $10^{19}$eV
(in fact it appears to harden),  possibly indicating a new hard component
in the cosmic-ray spectrum above the cutoff.

The most straightforward interpretation of the observation of cosmic
rays above the GZK cutoff is that the sources of the cosmic rays must
be ``local'' (within 50 Mpc or so).  Since cosmic rays above the GZK
cutoff should be largely unaffected by intergalactic and galactic
magnetic fields, there should be candidate cosmic accelerators
associated with the arrival direction of the UHE cosmic
rays.\footnote{In this paper, {\it ultra-high energy} (UHE) will refer
to energies near or above the GZK cutoff, which we take as $5 \times
10^{19}$eV.}  But there seems to be no unusual astrophysical sources
such as active galactic nuclei or quasars {\it within 50 Mpc} in the
direction of the events above the GZK cutoff.  There are hints of
unusual sources at cosmological distances (redshift of order unity)
associated with the arrival directions of the highest energy cosmic
rays \cite{elbert,glennys-peter}, but there is no way a UHE nucleon or
nucleus could reach us from such distances.  This suggests that new
physics or astrophysics is required to understand the nature,
production, composition, and propagation of such high-energy cosmic
rays.  This issue has attracted a lot of attention, because even the
most conservative explanations involve exciting new physics.

The first possible explanation is that the UHE cosmic rays actually do
originate within 50 Mpc from us but we cannot see the accelerator.
One might argue that the fact that there are events above the GZK
cutoff is evidence that there must be galactic or extragalactic
regions where the magnetic fields are far stronger than generally
believed, so the arrival direction would not be correlated with the
source direction \cite{colgate}.  Or, perhaps the reason we do not
observe any vestige of the cosmic accelerator is that the acceleration
results from a transient phenomenon such as a cosmic defect
\cite{cosmicstring} or a gamma-ray burst in a relatively nearby galaxy
\cite{gamma-ray-burst}.  In these scenarios the source could be within
50 Mpc and a primary cosmic-ray hadron could reach us without
appreciable energy loss.  Note however that the integrated flux
reduction factor for a $3.2 \times 10^{20}$ eV cosmic ray is of order
100 or more at 50 Mpc (c.f., Fig.\ 2 of Ref.\ \cite{elbert}).

A second possibility is that the UHE cosmic rays are not nucleons, and
originate from sources at cosmological distances without suffering
energy loss while propagating through the radiation backgrounds.
Stable non-hadronic primary candidates are electrons, photons, and
neutrinos.  Electrons are not candidates since their energy loss
during propagation is huge. 
Photons interact with the CMB and 
RBR
and lose energy by
$e^+e^-$ production, resulting in a propagation distance comparable to
protons.  Furthermore, the highest energy cosmic-ray events are
unlikely to be due to photons because photon-induced showers do not
resemble the observed Fly's Eye shower \cite{vasquez} and the muon
content of other events is consistent with expectations for a hadronic
rather than photon primary \cite{akeno}.  Neutrinos could reach us
from cosmological distance without energy loss, but it was shown that
even allowing for new physics, the observed events could not be due to
neutrino-induced showers \cite{birdman}.  Even with the meager number
of UHE events it seems very likely that the atmospheric shower was
induced by a strongly interacting particle.\footnote{Weiler has
proposed a hybrid model \cite{weiler-nu} where UHE neutrinos propagate
from cosmological distances, then produce hadrons in collisions with
massive neutrinos in the galactic halo.  This model seems to require
several parameters to be finely tuned.}

A third possibility is that the primary is a hadron, as suggested by
the properties of the observed air showers, but for some reason it has
a longer path length than a normal nucleon.  There are two ways this
might happen.  The first way is if the cross section for interaction
with the CMB is smaller.  The second way is for the energy threshold
for {\it resonant} photoproduction interactions to be higher, so that the
GZK bound is ``postponed" to higher energy.  Farrar \cite{glennysI} pointed
out that the existence in certain supersymmetric models of a new stable
hadron (e.g., $S^0$) with a mass of several GeV can realize this
possibility.  Due to its greater mass and its large mass-splitting to
its resonances, the threshold energy for resonance pion production is
increased compared to a proton.  This increases its effective path
length and postpones the GZK cutoff to higher energy.

A complete calculation of energy loss due to pion photoproduction,
redshift, and $e^+e^-$ production was done by Chung, Farrar and Kolb\cite{dan}
(CFK) for $S^0$'s in the mass range around 2 GeV.  Their analysis demonstrated
that the effective range through the CMB of a neutral primary as light as
2 GeV, with cross sections typical of a neutron but with a larger gap to the
lowest resonance, would be fifteen times more than a nucleon.  Although their
detailed analysis involved masses of around 2-3 GeV, they pointed out that
one might entertain the possibility that the primary is {\it much} more massive
than the nucleon.  Indeed, extensions of the standard model often involve
new colored particles with masses of order a TeV, which may be stable or
long-lived.  Since the pion production threshold energy is directly 
proportional to the mass of the primary, the GZK cutoff energy trivially
increases for a massive primary and, in addition, the fractional energy loss
per collision is smaller for a more massive primary.  Thus, from purely
kinematical considerations the path length of the primary increases
rapidly as its mass increases.  If a cosmic accelerator produces primary
protons with energies in excess of $10^{21}$ or so eV, then the accelerated
proton could collide with a proton at rest near the accelerator and produce
massive, high-energy particles with energy above the GZK cutoff: in
the collision of a primary of energy $10^{21}$eV with a proton at
rest, the invariant energy is $\sqrt{s} \sim10^3$TeV!

CFK coined the name ``uhecron" to describe a new species of long-lived
or stable, electrically neutral\footnote{The reason CFK take the
uhecron to be neutral is to avoid its interaction with galactic and
extragalactic magnetic field and the consequent energy loss due to
pair production and other mechanisms. This may not be an essential
requirement, depending on the distance to the accelerator.} hadron
which, due to its mass, can provide the answer to the cosmic-ray
conundrum.  Such a primary of mass $M$ can have sufficient range
through the cosmic backgrounds to reach us from sources at
cosmological distances ($L\sim 1\, {\rm Gpc}$) if its mass is more
than a few GeV, and its lifetime is greater than $10^6{\rm
s}(M/3{\rm GeV})(L/1{\rm Gpc})$.

But, as CFK pointed out, the fact that the uhecron is hadronic does
not guarantee that it will shower in the atmosphere like a nucleon.
If the uhecron contains a heavy constituent of mass $M$, the fraction
of its momentum which is carried by light degrees of freedom is
approximately $ \Lambda_{QCD}/M$.  It is energy from the light degrees
of freedom that is released in a typical hadronic interaction.  Thus,
if the uhecron contains a massive constituent, the energy spectrum of
the particles produced in a uhecron-nucleon interaction will be very
soft.  CFK argued that for a sufficiently massive uhecron it should be
possible to differentiate between a shower induced by a uhecron and a
shower induced by a proton.

The purpose of this paper is to determine the maximum possible mass of
a primary consistent with present observations by comparing the shower
of a massive uhecron with the development of the highest-energy
cosmic-ray shower observed, the $E\sim 3.2 \times 10^{20}$eV event
observed by the Fly's-Eye detector \cite{flys}.  On the basis of the
comparison, we show that the uhecron must be less massive than about
50 GeV in order to explain the highest energy event.
 
We also perform a more general comparison of proton-induced and
uhecron-induced air showers to find means of discriminating between
them.  Since the fluctuations in the development of hadronic showers
is large, it is impossible to identify reliably the difference between
a massive uhecron and a proton on a shower-by-shower basis.  But we
show that with sufficient statistics a uhecron with mass larger than
10 GeV may be distinguishable from proton showers.

It is important again to restate explicitly the picture of the uhecron
we will adopt for this study.  We will assume the uhecron consists of
a single constituent which accounts for the bulk of the mass of the
particle, surrounded by light hadronic degrees of freedom (gluons
and/or light quarks).  This definition of a uhecron strictly does not
encompass the archetypical uhecron, the $S^0$ \cite{glennysI, dan} which
is a bound state of light quarks and a relatively light gluino
($uds\tilde{g}$).  The momentum of the $S^0$ is shared roughly equally
by all constituents, so its shower properties are expected to resemble
those of a nucleon.\footnote{Furthermore, even if the $S^0$ momentum
were carried by a single constituent, our results (see below) indicate
that with a mass of order 2 GeV its shower is not distinguishable from
a proton.}  However there are a number of interesting uhecron
candidates for which the present definition is appropriate:
\begin{enumerate}

\item Raby has proposed a class of low-energy SUSY models in which the
gluino is the lightest supersymmetric particle \cite{raby}.
Presumably the lightest color-singlet state containing the gluino is
the $R^0$ (a gluino--gluon color singlet).  For gluinos more massive
than the QCD scale, the mass of the $R^0$ would be approximately the
mass of the gluino, which could be anywhere from a few GeV to 100 GeV
in his model.

\item Extensions of the standard model often include new colored
particles of mass in the 1 to 10 TeV range, often stable because of
some accidental symmetry or a new conserved quantum number (see
\cite{dan,rabi} for examples).  Note that cosmological restrictions
require them to be unstable over some mass ranges \cite{nardi-roulet}.

\item The uhecron may be a massive magnetic monopole (mass around
$10^{10\pm1}$GeV), accelerated by galactic magnetic fields to UHE
energies \cite{weilerpole}.  The magnetic monopole would have a small
core, but could also be surrounded by a gluon cloud.  Since the
monopole is accelerated by galactic fields, there need be no striking
astrophysical sources in the direction of the incoming UHE cosmic
rays.  Note, however, that it has been argued that their spectrum and arrival
direction disagrees with observation \cite{ubatuba}.

\end{enumerate}

In the next section we describe the longitudinal development of air
showers.  After a very brief description of the parameterization of
the development of the air shower, we describe the procedure used in the
simulation of the phenomenon.  We then discuss the modifications
necessary to study the development of uhecron-induced showers.  In
Section 3 we present the results of the uhecron simulations and
contrast the development of uhecron showers and proton showers.  We
next compare the uhecron results to the data for the highest-energy
event observed and derive a limit on the mass of the uhecron.  We then
discuss the kind of limit one might conceivably obtain with a larger
data set.  Finally, before ending the section, we describe the
ground-level particle content of a uhecron shower as might be observed
with a ground array of detectors.  We then present a short concluding
section.

\section{Ultra High Energy Air Showers}
\label{sec:uheas}

\subsection{Shower development}
\label{detectors}
\hspace*{1em}
The highest energy event was detected \cite{flys} by the Fly's Eye
collaboration using the atmospheric fluorescence technique
\cite{fluores}.  An air fluorescence detector measures the
longitudinal profile of the air shower.  This profile reflects the
development of the cascade generated by the interaction of the primary
particle with the atmosphere.

The number of particles in the shower grows until it reaches a maximum
($N_{MAX}$) at a certain atmosphere depth ($X_{MAX}$). The
longitudinal development of the number of particles $N$ as a function
of atmospheric depth $X$ (measured in g\,cm$^{-2}$) in the shower can
be fit by the function \cite{cp}:\footnote{This function is a slightly
better fit than the familiar Gaisser--Hillas function \cite{clem} and
was found empirically\cite{cp}.}
\begin{equation}
\frac{N(X)}{N_{MAX}} =  \left( \frac{X-X_0}{X_{MAX}-X_0} \right)
^{(X_{MAX}-X)(10+0.02X)^{-1}} \ ,
\label{eq:cp}
\end{equation}
where $X_0$, $N_{MAX}$, and $X_{MAX}$ are fit parameters.  The energy
of the event is proportional to the integral of $N(X)$, the total
number of charged particles in the shower.

In addition to giving the energy of the primary, the longitudinal
development of the shower can be used to constrain the nature of the
primary.  Although there are large fluctuations in $X_{MAX}$ on a
shower-to-shower basis, the $X_{MAX}$ distribution for many showers
differs depending on whether the primary particle is a proton or a
nucleus.  The reason for this is that a nucleus showers like an
ensemble of lower energy nucleons, so that for a given total primary
energy, the mean $X_{MAX}$ for the distribution is larger for a proton
shower than for an iron shower, i.e., the proton shower peaks deeper
in the atmosphere.  However, due to the large fluctuations in
$X_{MAX}$ the composition determination has a large uncertainty unless
there are a large number of events.  Thus, there is no clear
indication of the composition for the very highest energy cosmic rays
\cite{Watson97}.
 
Most of the statistics of UHE events have been gathered by extensive
air shower (EAS) arrays rather than with the air fluorescence
technique.  EAS arrays measure the density of particles as a function
of the distance from the core, from which the energy of the primary is
inferred. The ratio of muons to the electromagnetic density at some
distance from the core of the shower (600\,m or more) can be used to
tell whether the showers were induced by photons or hadrons.
Discrimination between nucleons and nuclei is more difficult but
feasible with sufficient statistics.

In this work, we simulate the development of showers initiated by a
320 EeV uhecron of various masses between 2.5 and 100 GeV.  We compare
the results of their longitudinal development to the shower observed
in the highest energy event. We also estimate the ratio between muon
and total charged particle densities at ground level.  We find that it
is not possible to distinguish a proton from a uhecron on a
shower-to-shower basis, unless the uhecron is extremely massive. However,
with sufficient statistics it may be possible to discern the difference
between the showers of a uhecron of mass greater than about 10 GeV and those
of a nucleon. 

We emphasize that our goal is to look at {\it differences} between the
shower development of ultrarelativistic nucleon and uhecron primaries.
It is known from new data at lower energies that existing shower simulations
are inadequate in several respects.  This new data will allow improvements
in the modeling of the QCD processes, just as modeling of hadronic collisions
relevant for discovering new particle physics at LEP, SLC and the FNAL collider
gradually improved with more complete data on conventional processes.
Likewise, by the time enough events have been accumulated above the GZK bound to
apply the kind of analysis we envision, there will be a vastly larger number
of events at energies below the GZK bound, where we can be confident
the primaries are dominantly conventional particles.  That will allow the
understanding of conventional particle showers to be improved to the
point that we can have confidence in the modeling of the showers due to
proton primaries above the GZK bound.  Then a deviation due to new
particles can be studied.  Our purpose below is to estimate the extent
of the {\it deviation} in showering between proton and uhecron
primaries which follows from the difference in the momentum-fraction
carried by light consituents in a uhecron as compared to a nucleon.
We expect this to be a robust discriminator, very weakly dependent of the
details of how the light constitutents produce a shower.

\subsection{Air Shower Simulation}
\hspace*{1em}
An air shower simulation requires an event generator to simulate the
interaction of each shower particle with an atmospheric nucleus and a
simulation of the cascade development. In our simulations we use Aires
(AIR Shower Extended Simulations) \cite{aires} as the cascade
simulator\footnote{The cascade development used in Aires is similar to
the one used in MOCCA \cite{MOCCA}.  For a comparison between Aires
and MOCCA, see \cite{COMP}.} and Sibyll \cite{sibyll} as the event
generator.  We modified Aires and Sibyll to include a new particle, the
uhecron. The modification to Aires was straightforward: adding
a new particle to the cascade development.  The modifications to
Sibyll were more extensive.

Sibyll uses a combination of a model for low energy hadron-hadron
interactions, a model for the ``hard'' part of the cross section, and
a model to go from hadron-hadron to hadron-nucleus interactions.  For
the initial very high energy interactions at $\sqrt{s}$ up to 800 TeV,
an extrapolation from data taken in laboratory experiments is
required.  In Sibyll this is done with the Dual Parton Model augmented
by minijet production (see \cite{sibyll} and references therein) as
will be briefly described below.

In the low energy regime ($\sqrt{s}\simeq$10 to 20 GeV or
$E_{\rm{lab}}= $50 to 200 GeV) the observed hadronic cross section
exhibits Feynman scaling.  At these energies, hadron-hadron
interactions can be represented by the production and fragmentation of
two QCD strings (see \cite{sibyll} and references therein). The
collision occurs between the incoming hadron (either a baryon or a
meson) and nucleons in a target nucleus, which is randomly taken to be
oxygen or nitrogen. In the case of a baryon-baryon collision the
energy of each baryon is divided between one quark $q$ and one diquark
$qq$.  Sibyll employs a structure function to describe the fraction
$x$ of the baryon energy carried by the quark fragment given by
\begin{equation}
f_{q}(x) =  \frac{(1-x)^{\alpha}}{(x^{2}+\mu^{2}/s)^{1/4}}\  .
\label{eq:struc}
\end{equation}
Here $\mu$ is considered as an ``effective quark mass" and is taken to
be 0.35 GeV and $\alpha$ is taken to be 3.0.  The diquark carries the
remaining energy of the hadron, aside from that part of the hadron
energy going into minijet production.

QCD strings are formed between the quark of the incoming particle and
the diquark of the target, and vice-versa. The formation of strings
obeys energy and momentum conservation. At the end of this process,
hadronization occurs.  The energy of each produced particle is
generated according to the Lund fragmentation function \cite{sibyll}.
At each interaction, ``leading particles'' are produced. They are the
baryons which contain the original diquark from the incoming particle
and from the target nucleons. The energy carried by these particles is
described by a harder fragmentation function than that for the
non-leading particles.

\subsection{Modifications to Sibyll to include a uhecron}
\label{subsec:uhe}
\hspace*{1em}
We modify Sibyll as described below to account for the difference
between the energy deposition properties of a hadron containing a
heavy constituent ``clothed" by light quarks and gluons (a uhecron)
and the energy deposition of a normal hadron whose momentum is carried
only by light constituents.  As is well understood in the context of Heavy
Quark Effective Theory (HQET) \cite{hqet}, the interaction of the uhecron
with matter is dominated by the interaction of its light degrees of freedom
(quarks and/or gluons).  

Hadronic total cross sections are determined by the spatial extent of the
colored quanta.  By the uncertainty principle, this is greatest for the
lightest degrees of freedom, so we take as our canonical choice for the
uhecron-nucleon cross section $\sigma_{UN} = \sigma _{\pi N}$, since
the light degrees of freedom are light quarks, antiquarks, and gluons
for nucleon, mesons and uhecron.  However we also consider explore the effect
of taking $\sigma_{UN} = 1/10 \sigma _{\pi N}$. This lower cross section
is motivated by lattice 
QCD studies which show that the radii of glueballs tend to be two to four
times smaller than the radii of ordinary mesons.  The reduced size of a hadron
containing a valence color-octet particle is understood as being due to its
larger color charge and thus stronger confining potential.  Thus, if the
uhecron contains a heavy color-octet constituent such as a gluino, the lower
cross section may be appropriate. 

In an ultrarelativistic uhecron, as in a B meson, the heavy constituent (mass
$m_Q$) has most of the momentum of the uhecron and the light degrees
of freedom are left with only a small fraction. In our modification of Sibyll,
we use for the uhecron the standard 
Peterson function \cite{prd} to describe the fraction of energy $z$ of
a heavy hadron carried by its heavy quark $Q$:
\begin{equation}
f_Q(z) = \frac{1}{z} \left[ 1 - \frac{1}{z} - \frac{\epsilon_Q}{1-z} 
\right]^{-2}.
\label{eq:pet}
\end{equation}
Here $\epsilon_{Q}$ is proportional to $\Lambda_{\rm
QCD}^{2}/m_{Q}^{2}$.  We use this function to replace the hadronic
structure function, Eq.\ (\ref{eq:struc}), and the fragmentation
function of a standard hadron.  In the fragmentation function the
final hadron carries a fraction $z$ of the energy of its parent
``quark".  This is a hard fragmentation function, and in the limit
$m_{Q} \rightarrow \infty$ it approaches a delta function peaked at
1. We use the Peterson fragmention function since it is in good
agreement \cite{prd} with data for $b$ quark interactions and has the
required qualitative behavior in the limit of extremely heavy
constituent.

In the hadron-hadron interaction model employed by Sibyll the quarks
that belonged to the original baryon undergo a hadronization process
in which different probabilities are taken for forming various hadrons
\cite{sibyll}. In the uhecron case, the only possibility is for the heavy
``quark" $Q$ to hadronize as a uhecron.

Sibyll also includes diffractive dissociation. When diffraction occurs
the incoming particle is excited into a higher mass state and then decays 
if its mass is not large.  For larger masses, as in the uhecron case, the
excited state does not decay and is instead split, e.g., into a quark
and a diquark in the baryon case.  These move apart stretching a
string between them.  The string fragments as in the nondiffractive
case.  
The only modification in the diffractive part of Sibyll was to modify
the minimal mass limit of the excited state according to the uhecron
mass.

A final modification was related to the ``hard'' part of the cross
section in which the momentum transfer is large. In this case Sibyll
models the production of minijets with energies of several GeV.  As
the uhecron interacts always with low energy because most of the
initial energy is carried by the heavy quark that does not interact,
we neglect minijet production when the incident particle is a
uhecron. Of course, ordinary hadrons are created even in a
uhecron-initiated shower and minijets can be produced from the
subsequent interactions of these normal hadrons.

\newpage

\section{Results of the simulations}
\label{sec:res}

\begin{figure}[!p]
\centering 
\leavevmode \epsfxsize=225pt \epsfbox{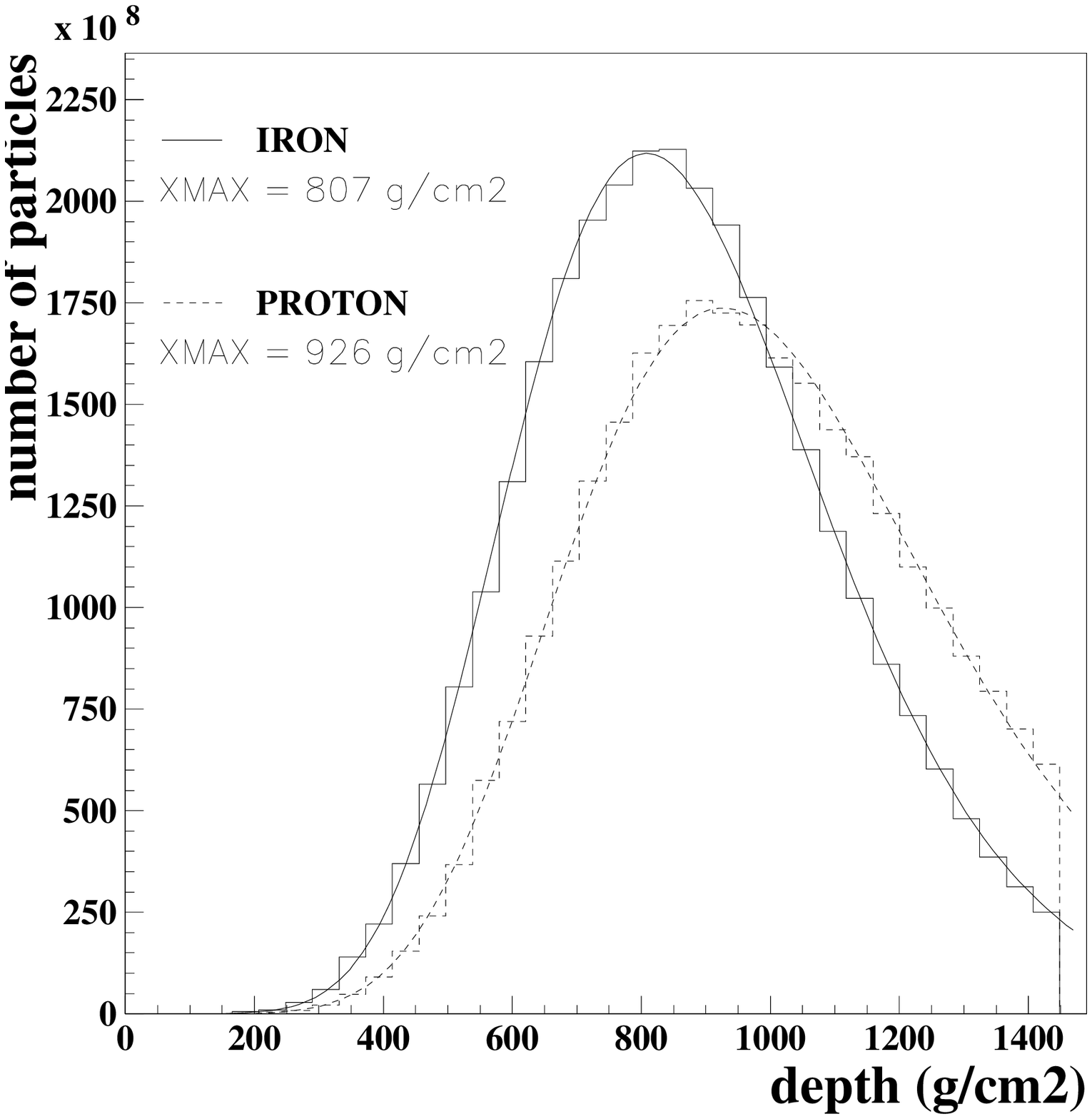}
\epsfxsize=225pt \epsfbox{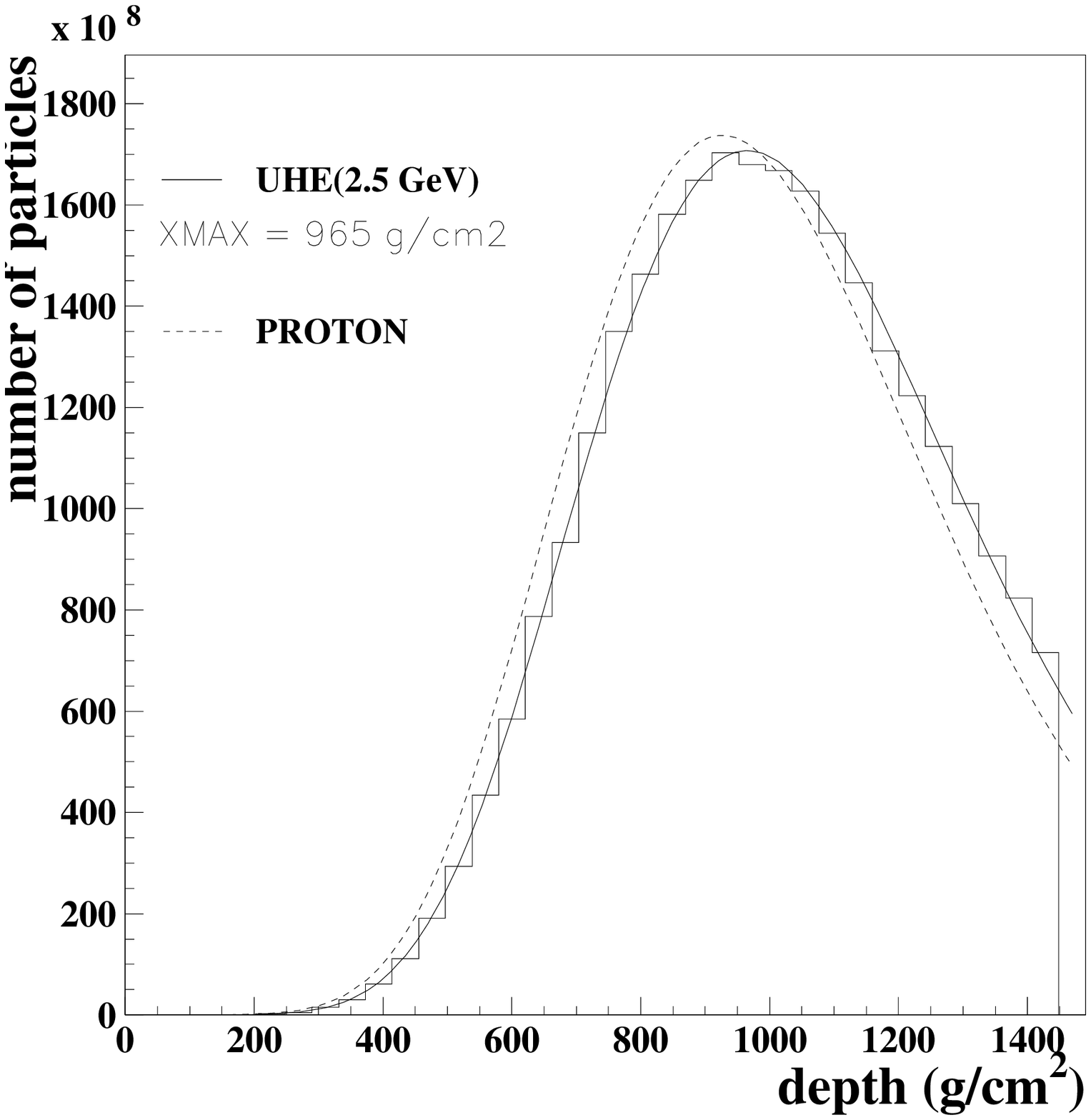} \\
\vspace*{24pt}
\leavevmode \epsfxsize=225pt \epsfbox{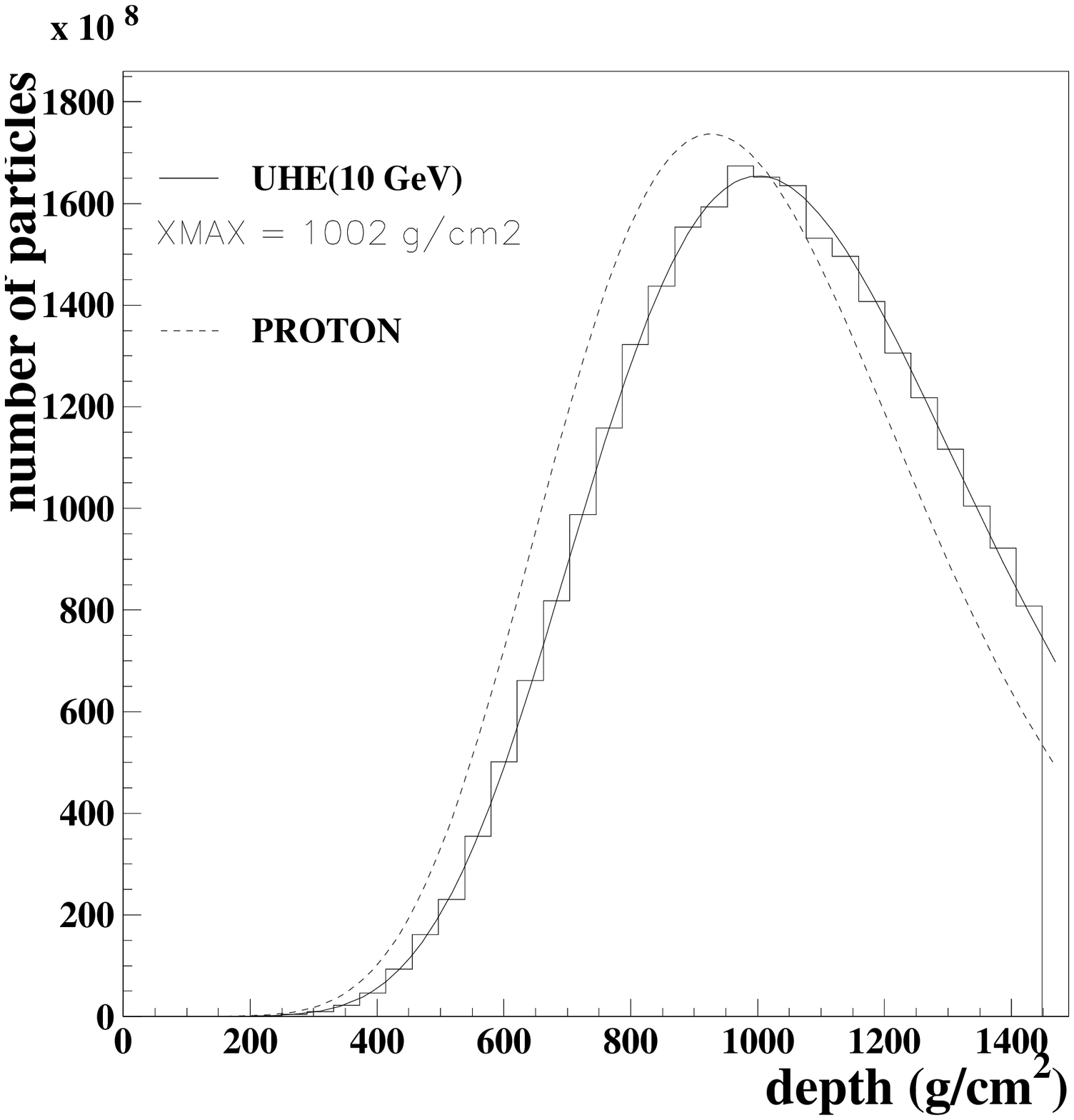} 
\epsfxsize=225pt \epsfbox{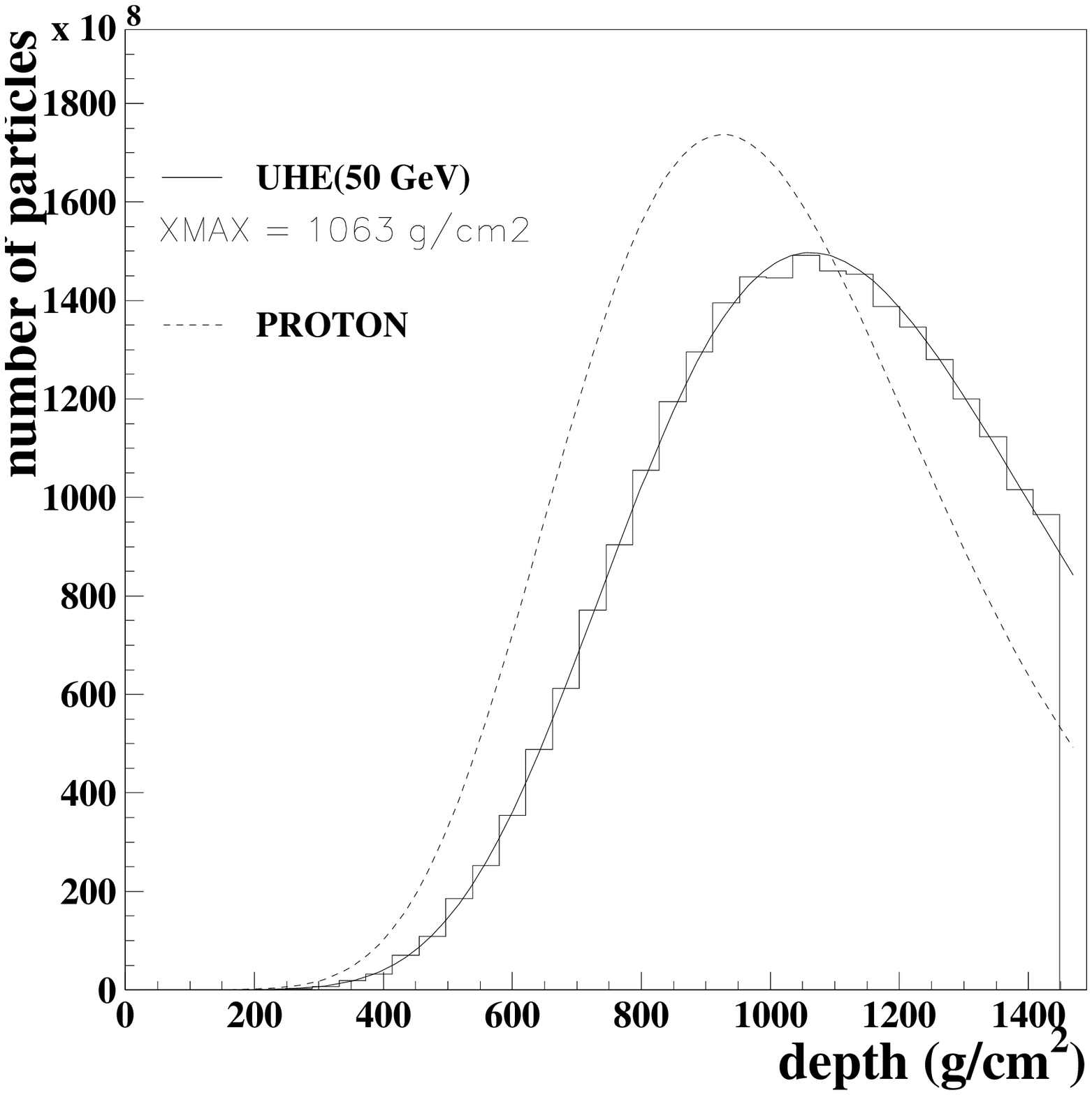}
\caption[fig1]{\label{fig:prof} Longitudinal development for 320 EeV
showers based on 500 showers for each. All plots are compared with 
the proton profile: a) iron; b) uhecron (2.5 GeV); c) uhecron (10 GeV); 
d) uhecron (50 GeV). The fitted $X_{MAX}$ is given for each primary. }
\end{figure}

\begin{figure}[!p]
\centering 
\leavevmode \epsfxsize=225pt \epsfbox{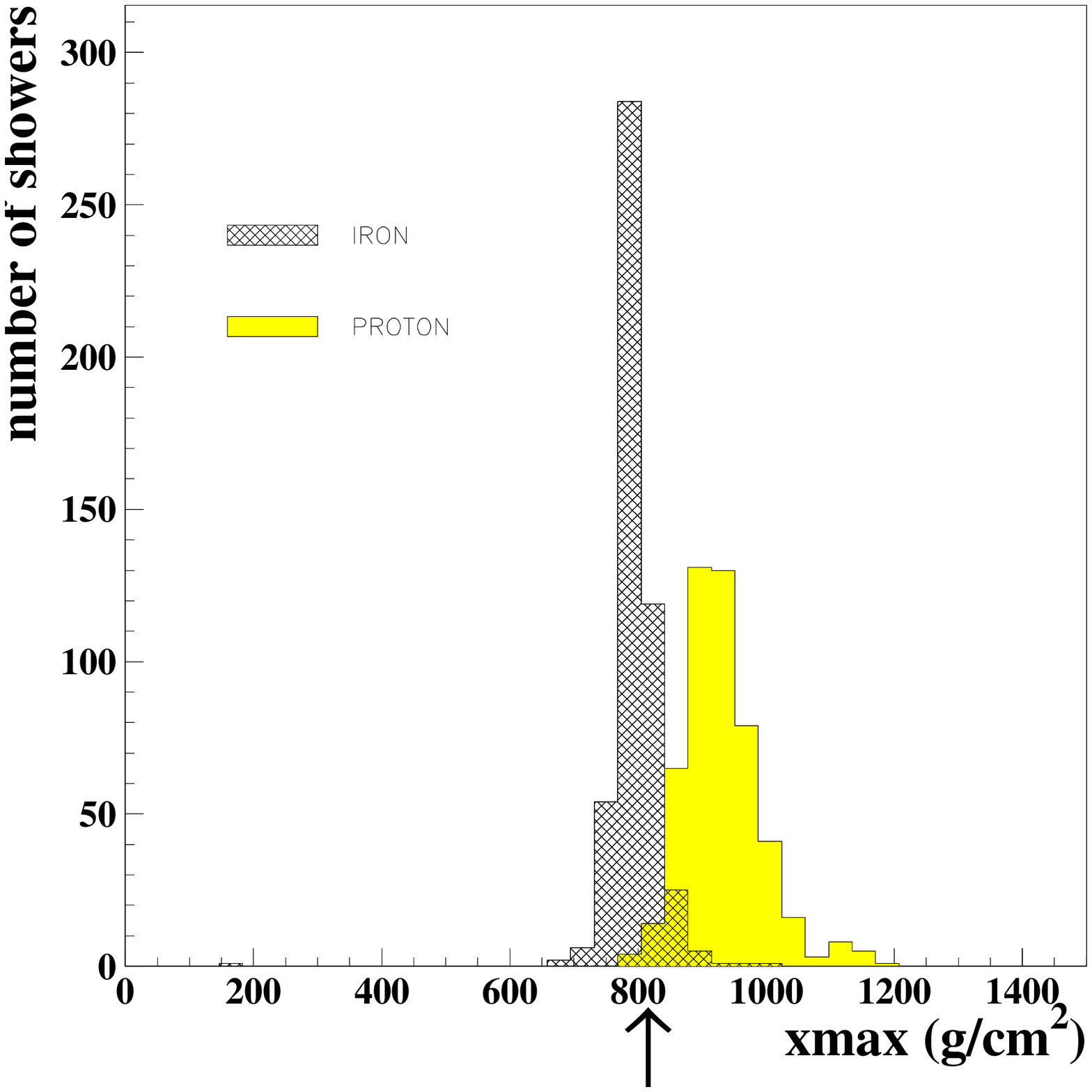}
\epsfxsize=225pt \epsfbox{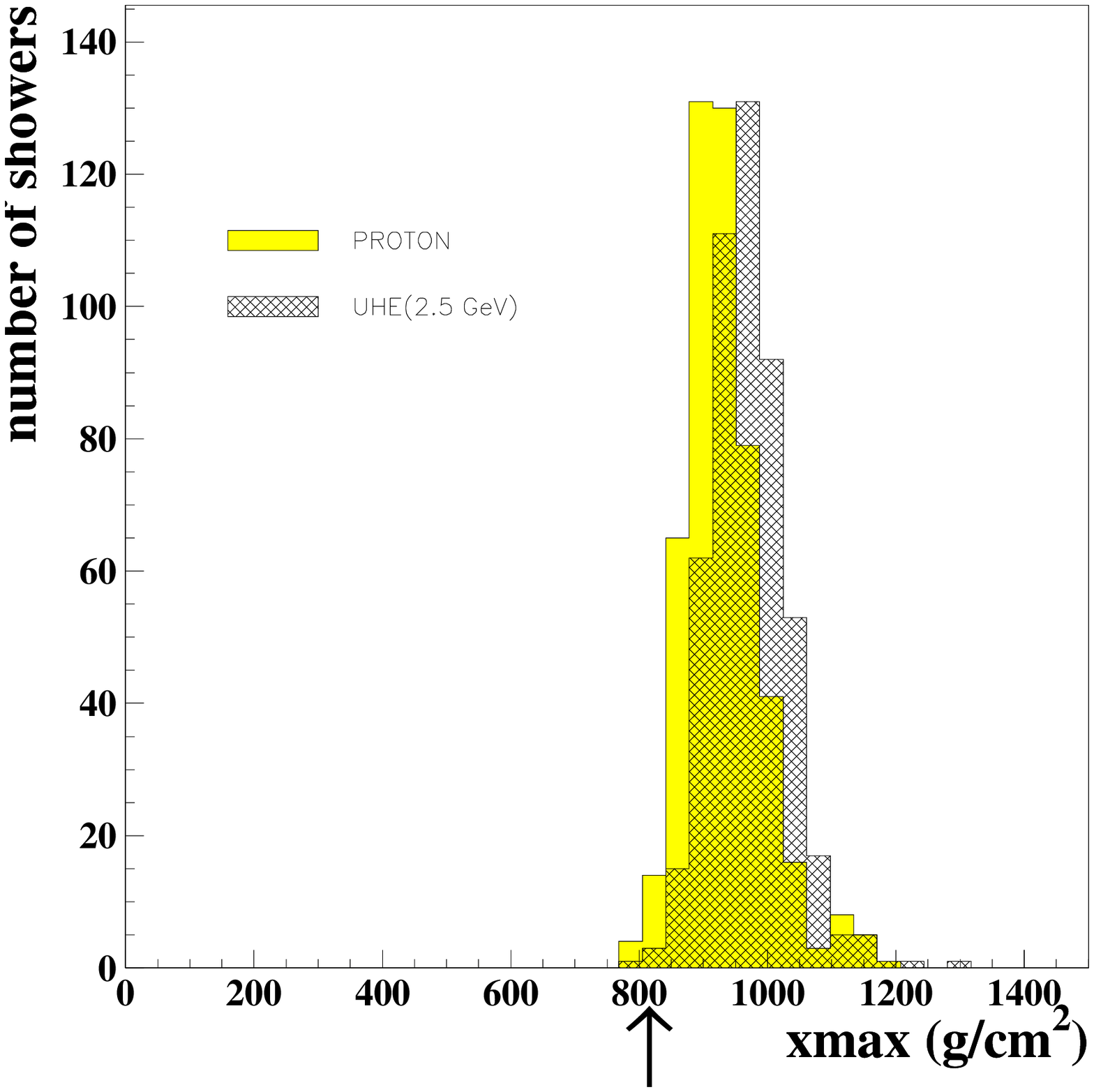} \\
\leavevmode \epsfxsize=225pt \epsfbox{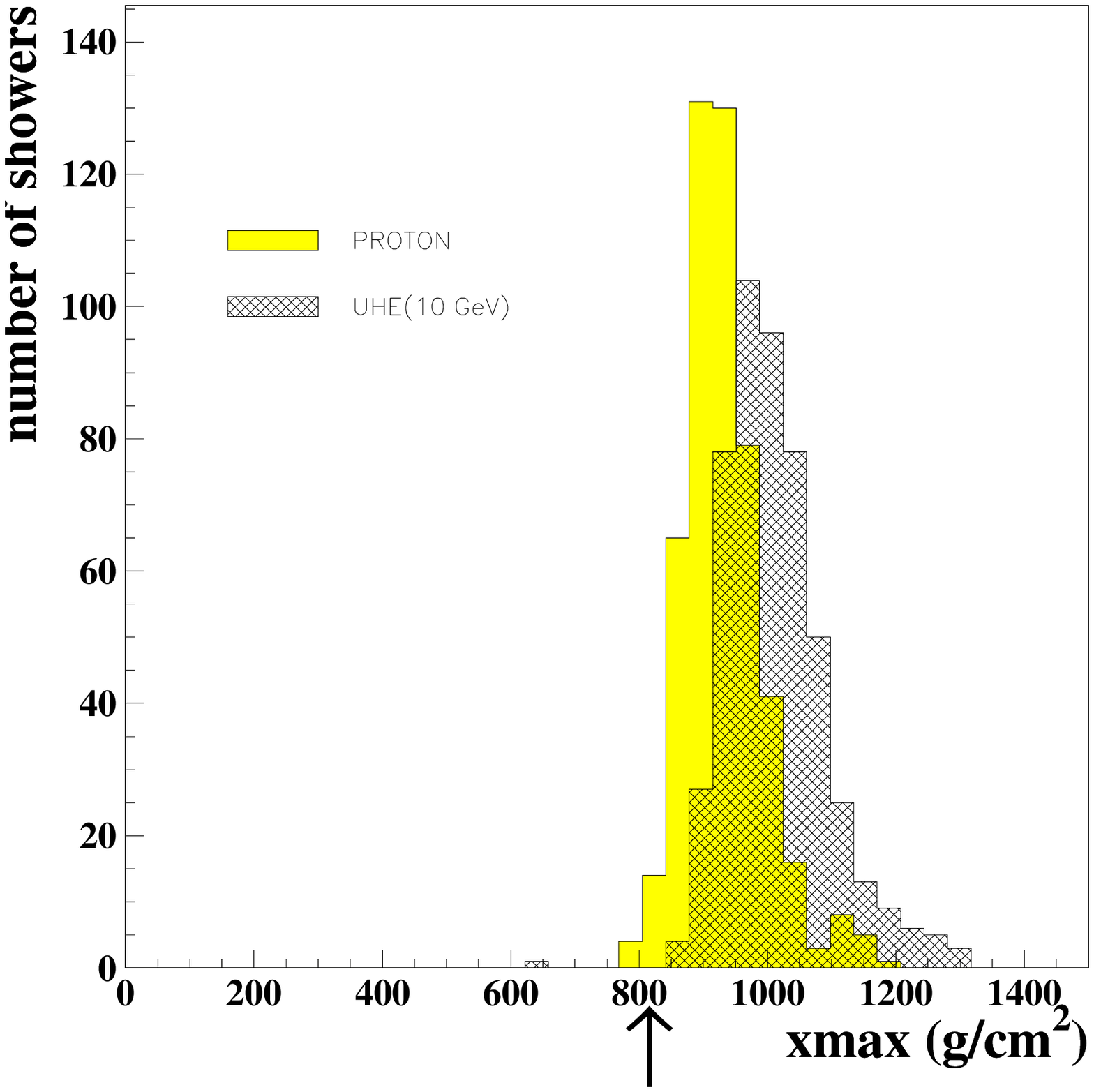} 
 \epsfxsize=225pt \epsfbox{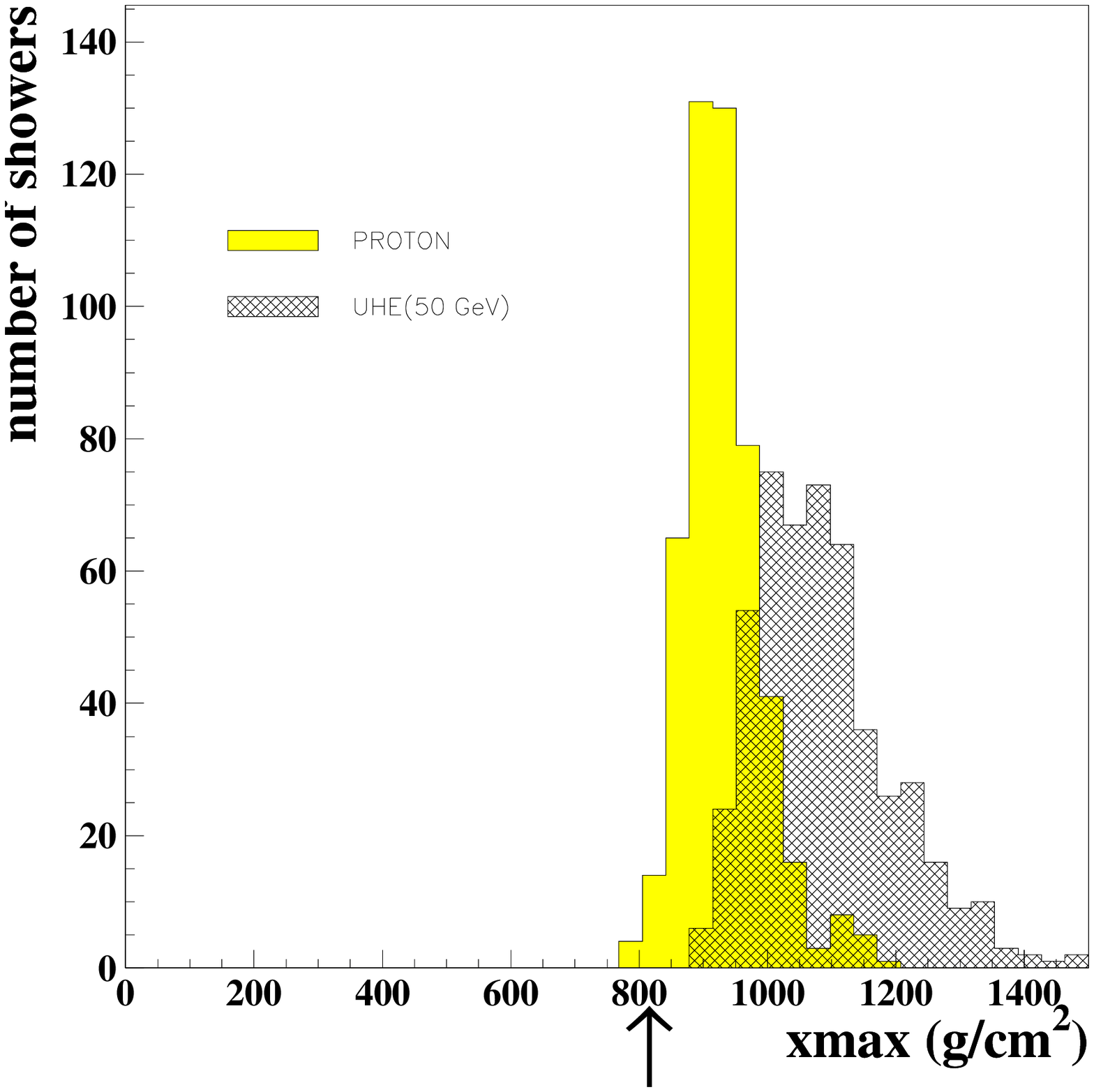}
\caption[fig2]{\label{fig:xmax} $X_{MAX}$ distribution for 320 EeV
showers.  All plots are compared with the proton distribution: a)
iron; b) 2.5 GeV uhecron; c) 10 GeV uhecron; d) 50 GeV uhecron. The
distributions are based on 500 showers for each case.The arrows
indicate the value of $X_{MAX}$ for the Fly's Eye event,
$815^{+60}_{-53}$.}
\end{figure}

\subsection{Longitudinal shower development}
\hspace*{1em}
Assuming a shower energy of 320 EeV, in Fig.\ \ref{fig:prof} we show
the longitudinal distributions (see Section~\ref{sec:uheas}) for
protons, 2.5 GeV uhecrons, 10 GeV uhecrons, and 50 GeV
uhecrons.\footnote{Unless otherwise specified the uhecron-nucleon
cross section is the same as for a pion-nucleon interaction.  We
return to the reduced cross section case at the end of this section.}
For each case, the histogram is the single best representation of 500
shower simulations, and the solid line is the best fit to the
histogram using Eq.\ (\ref{eq:cp}).

The first plot shows iron and proton showers for a reference.  The
other plots in this figure show that $\langle X_{MAX} \rangle$
increases as the uhecron mass increases.  The shower maximum is deeper
for a heavier uhecron because the energy loss per uhecron--nucleon
collision decreases as the uhecron mass increases. The profile for a
typical 2.5 GeV uhecron resembles that of a proton, but $\langle
X_{MAX}\rangle$ is approximately 40 g\,cm$^{-2}$ deeper and $N_{MAX}$
is lower. The average profile for a 50 GeV uhecron appears very
different than the proton.  However, it would be hard to distinguish a
50 GeV uhecron from a proton for an individual shower just based on
its longitudinal distribution.  This is because of the large
fluctuations in the the value of $X_{MAX}$ for hadron-initiated
showers.

In order to study the $X_{MAX}$ fluctuations, we show in Fig.\
\ref{fig:xmax} the $X_{MAX}$ distribution for the 500 showers used in
Fig.\ \ref{fig:prof}. It is evident that as the uhecron mass increases
$\langle X_{MAX} \rangle$ increases and the distribution becomes
broader.  This broadening of the distribution makes it difficult for a
single event identification of a high-mass uhecron since $X_{MAX}$
fluctuates more when the particle mass is higher.  Conversely, given a
large number of events, the observed dispersion in $X_{MAX}$ would
itself be a useful tool in identifying a uhecron and inferring its
mass.

\subsection{Uhecron showers compared to observation}
\label{compatibility}
\hspace*{1em}
{}From the $X_{MAX}$ distributions shown in Fig.\ \ref{fig:xmax} one
can estimate the compatibility between these simulated showers and the
Fly's Eye event.  The energy of the observed event \cite{flys} is
$320^{+92}_{-94}$ EeV, the $X_{MAX}$ is $815^{+60}_{-53}$
g\,cm$^{-2}$, and the zenith angle is
$43^\circ\!\!.9^{+1^\circ\!\!\!.8}_{-1^\circ\!\!\!.3}$ (the
uncertainty is a combination of statistical and systematic
effects). We use these values for the primary energy and the zenith
angle in our simulations.

We compare our results with this Fly's Eye event because it is the
only one with energy above the GZK cutoff detected with the
atmospheric fluorescence technique. This allows a comparison of
$X_{MAX}$ and the longitudinal distribution of our simulations with
the observed event.

One can get a qualitative understanding why there will be a mass limit
for the uhecron by examining Fig.\ \ref{fig:xmax}.  The distribution
for a 2.5 GeV uhecron is not very different from the proton
distribution\footnote{Nor have we included sufficient refinement in the
modeling for such a small difference to be significant.}.  But as the uhecron
mass increases, the distribution becomes distinguishable from the proton
distribution.  The trend is that as the mass increases, $\langle X_{MAX}
\rangle$ increases, as does the variance of the distribution.

The 320 EeV event observed by Fly's Eye had $X_{MAX}$ of 815
g\,cm$^{-2}$.  This value is quite low compared to the mean of the
distributions for large uhecron mass and therefore the observed event is
unlikely to come from the probability distribution for large uhecron masses.
The fact that the width of the uhecron's $X_{MAX}$ distribution increases
with mass complicates this simple picture, but one can still estimate the
maximum uhecron mass consistent with observation, keeping in mind the fact
that there is but one event.

We will assume that the observed event was generated by a uhecron of
unknown mass.  The probability that the event resulted from a uhecron
of mass $M_i$ is proportional to the probability for a uhecron of mass
$M_i$ to generate the observed event (Bayes' rule assuming equal
priors).  The probability distributions in Fig.\ \ref{fig:xmax} can be very
well fit by a Gaussian plus an exponential tail.  Let us refer to the
normalized $X_{MAX}$ probability distribution for a uhecron of mass
$M_i$ as $P(X_{MAX};M_i)$. The value of $P(815\,{\rm g\,cm}^{-2};M_i)$
is proportional to the probability that the observed event was drawn
from $P(X_{MAX};M_i)$.  We then compare the probabilities for various
choices for the uhecron mass.  

\begin{table}
\caption{\label{bayes} The relative probability that a uhecron of mass $M_i$
produces an event with the indicated value of $X_{MAX}$.}
\vspace{12pt}
\centering
\begin{tabular}{ccc}
\hline \hline
\hspace*{18pt} Energy \hspace*{18pt} & 
\hspace*{18pt} $\Frac{P(815\,{\rm g\,cm}^{-2};M_i)}
                     {P(815\,{\rm g\,cm}^{-2};2.5~{\rm GeV})}$
  						\hspace*{18pt} &
\hspace*{18pt} $\Frac{P(875\,{\rm g\,cm}^{-2};M_i)}
                     {P(875\,{\rm g\,cm}^{-2};2.5~{\rm GeV})}$    
                                                \hspace*{18pt} \\ \hline 
2.5 GeV & 1.00 & 1.00 \\
5 GeV & 0.23 & 0.39\\
10 GeV & 0.24 & 0.33\\
20 GeV & 0.17 & 0.21\\
40 GeV & 0.08 & 0.09 \\
50 GeV & 0.11 & 0.12 \\
100 GeV & 0.04 & 0.06 \\
\hline\hline
\end{tabular}
\vspace{4mm}
\end{table}

We arrive at a limit on the mass of the uhecron by demanding that
$P(815\,{\rm g\,cm}^{-2};M_i)$ is at least 10\% of its peak value.
Since $P(X_{MAX};M_i)$ varies only slowly between $M_i = m_p$ and 2.5
GeV, we conservatively use $P(815\,{\rm g\,cm}^{-2};2.5 ~{\rm GeV})$ as our
reference value for determining the uhecron mass limit\footnote{By this
procedure, we remove our sensitivity to shower-modeling inadequacies
which may produce a shift between the predicted and observed $X_{MAX}$
distributions at lower energies.}. In Table \ref{bayes} we show the ratio $P(815\,{\rm g\,cm}^{-2};M_i)/P(815\,{\rm
g\,cm}^{-2};2.5\  {\rm GeV})$ for several values of $M_i$.  One sees that
for $M_i>40$ to 50\,GeV the probability is less than 10\% of its value at
$M_i = 2.5$ GeV. We therefore take this value as the largest uhecron mass
compatible with observation.\footnote{We also used $P(875\,{\rm g\,cm}^{-2};M_i)$,
where 875 g cm$^{-2}$ is the 1 $\sigma$ upper limit of the observed
event, and found that our limit does not change significantly.}
However, it must be noted that because large-mass uhecrons have
broader distributions, $P(815\,{\rm g\,cm}^{-2};M_i)$ does not
decrease as rapidly with $M_i$ as one might expect just on the basis
of the increase of $\langle X_{MAX} \rangle$.  For instance,
$P(815\,{\rm g\,cm}^{-2};100\,{\rm GeV})$ is only about half as large as
$P(815\,{\rm g\,cm}^{-2};50\,{\rm GeV})$.  The moral is that more data
is needed, but just on the basis of one event, a uhecron of mass
larger than 40 to 50 GeV seems very unlikely.

\subsection{Achievable mass limit with additional data}
\hspace*{1em}
Now we turn to the question of how one might improve the limit on the
uhecron mass with the observation of many more events.  This discussion 
is meant to be just an indication of what might be obtained.

The most distinctive feature of the $X_{MAX}$ distribution for uhecron
showers is the prominence of the tail for large $X_{MAX}$.  As the
mass increases, the tail becomes more pronounced.  If sufficient
events are observed to form the $X_{MAX}$ probability distribution
function, then one can test for the size of the tail.

\begin{figure}[!t]
\centering 
\leavevmode \epsfxsize=300pt \epsfbox{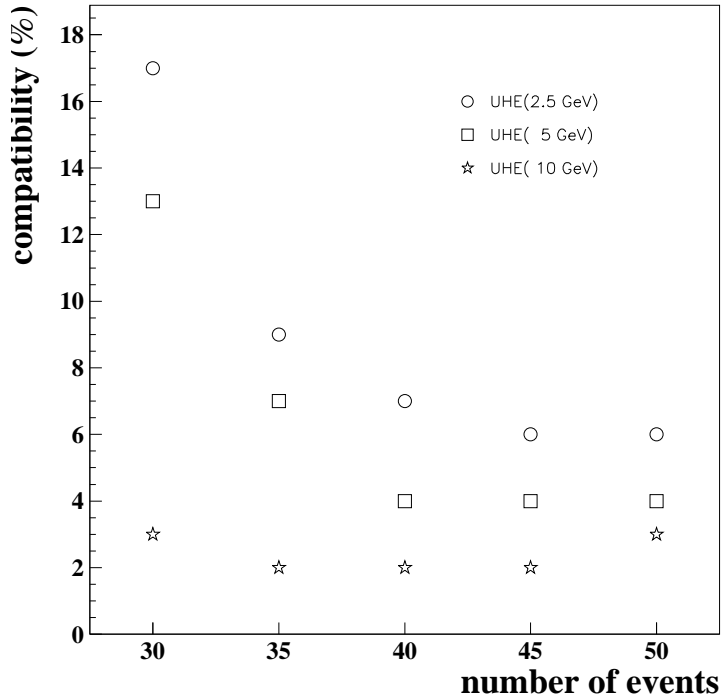}
\caption[fig3]{\label{fig:percent} Percentage of the time a number of
uhecron showers {\it cannot} be distinguished from a proton shower,
for several different choices of the uhecron mass.  For instance,
given a data set with 40 showers, there is a 7\% probability that the
2.5 GeV uhecron $X_{MAX}$ distribution function will be well fit by
the proton $X_{MAX}$ distribution function.  For a 5 GeV (10 GeV)
uhecron, the probability is 4\% (2\%) if one has data from 40
showers.}
\end{figure}

We proceed as follows:
\begin{enumerate}
\item For the proton and for various mass uhecrons, we fit the
probability distribution functions of Fig.\ \ref{fig:xmax} to a
Gaussian plus an exponential tail for large $X_{MAX}$.  (The
distributions are very well fit by this form.)
\item We generate a synthetic data set of $N$ events from the
probability distribution function for a uhecron of mass $M$.
\item We fit the $N$ events generated in this way with the proton
probability distribution function generated in step 1.  If it has a
reasonable $\chi^2$, then we proceed, if not, we say that the data set
generated in step 2 is incompatible with arising from proton-induced
showers.
\item We subject data sets with a reasonable $\chi^2$ to another test by
computing the number of events in the exponential tail and
comparing it to the number of events in the Gaussian part of the
distribution.
\item If this ratio is the same (within $\pm 30$\%) of that expected
from the proton distribution, we say that the generated events are
consistent with arising from the proton distribution function.
\item Then we repeat steps 2 through 5 many times.  Some data sets of
$N$ events will be consistent with a proton distribution, and some
will not be consistent.  We compute the percent of data sets that are
consistent with the proton.  If a large percentage of synthetic data
sets are consistent, then we are unlikely on the basis of $N$ events
to be able to differentiate between a uhecron of mass $M$ and a
proton.  If, on the other hand, only a small percentage of the
synthetic data sets are consistent with a proton distribution, then we
are very likely to be able to discriminate between a uhecron of mass
$M$ and a proton on the basis of $N$ events.
\item We repeat the procedure for different $M$ and $N$.
\end{enumerate}
The result of this procedure is shown in Fig.\ \ref{fig:percent}.  
The trends in Fig.\ \ref{fig:percent} are easy to understand.  For a
data set consisting of a given number of events, it is less likely for
a large-mass uhecron to be consistent with a proton distribution.  For
a given uhecron mass, as the number of events increases, it becomes
easier to differentiate a proton from a uhecron.  The apparent rise at
50 events for a 10 GeV uhecron is a statistical fluctuation.

We derive a ``potential'' mass limit from the fact that if the uhecron
mass is larger than 10 GeV, there is better than a 95\% probability
one would be able to tell that the $X_{MAX}$ distribution was not
generated by a proton from a data set with more than 30 events.

Again, we stress that this is just an illustrative calculation.
For instance, we have not included any smearing of the $X_{MAX}$ 
probability distribution functions as a result of finite detector resolution.
A full analysis of this issue must be left to the detector
collaborations, but we can get a preliminary idea of its importance as
follows. Let us assume that the experimentally determined
$X_{MAX}$ is Gaussian distributed about the actual $X_{MAX}$, with width
$\sigma_E \simeq 20$  g cm$^{-2}$.  As can be seen from Fig. 2, the proton
$X_{MAX}$ probability distribution function is also reasonably well fit by
a normal distribution, due to the stochastic nature of particle production
in hadronic showers.  The width of the intrinsic $X_{MAX}$ spreading is rather
large, $\sigma_I\simeq 50$ g cm$^{-2}$.  Thus the reconstructed $X_{MAX}$
distribution will be Gaussian with width $\sigma=\sqrt{\sigma_I^2+\sigma_E^2}$.
So long as $\sigma_E$ is smaller than $\sigma_I$, the experimentally measured
width of the distribution would not be much different than shown in Fig.\
\ref{fig:xmax} using only the intrinsic smearing.  A more serious issue would
be if there are non-Gaussian  tails in the detector resolution  at large
$X_{MAX}$. This would cause the observed proton distribution to resemble
a uhecron distribution. A true limit on the uhecron mass will require a
simulation including the detector response function, but unless there are
large non-Gaussian tails in the detector response, our estimates should give
a reasonable indication of what is achievable.   

\subsection{The effect of the uhecron cross section}
\begin{figure}[!t]
\centering 
\leavevmode \epsfxsize=225pt \epsfbox{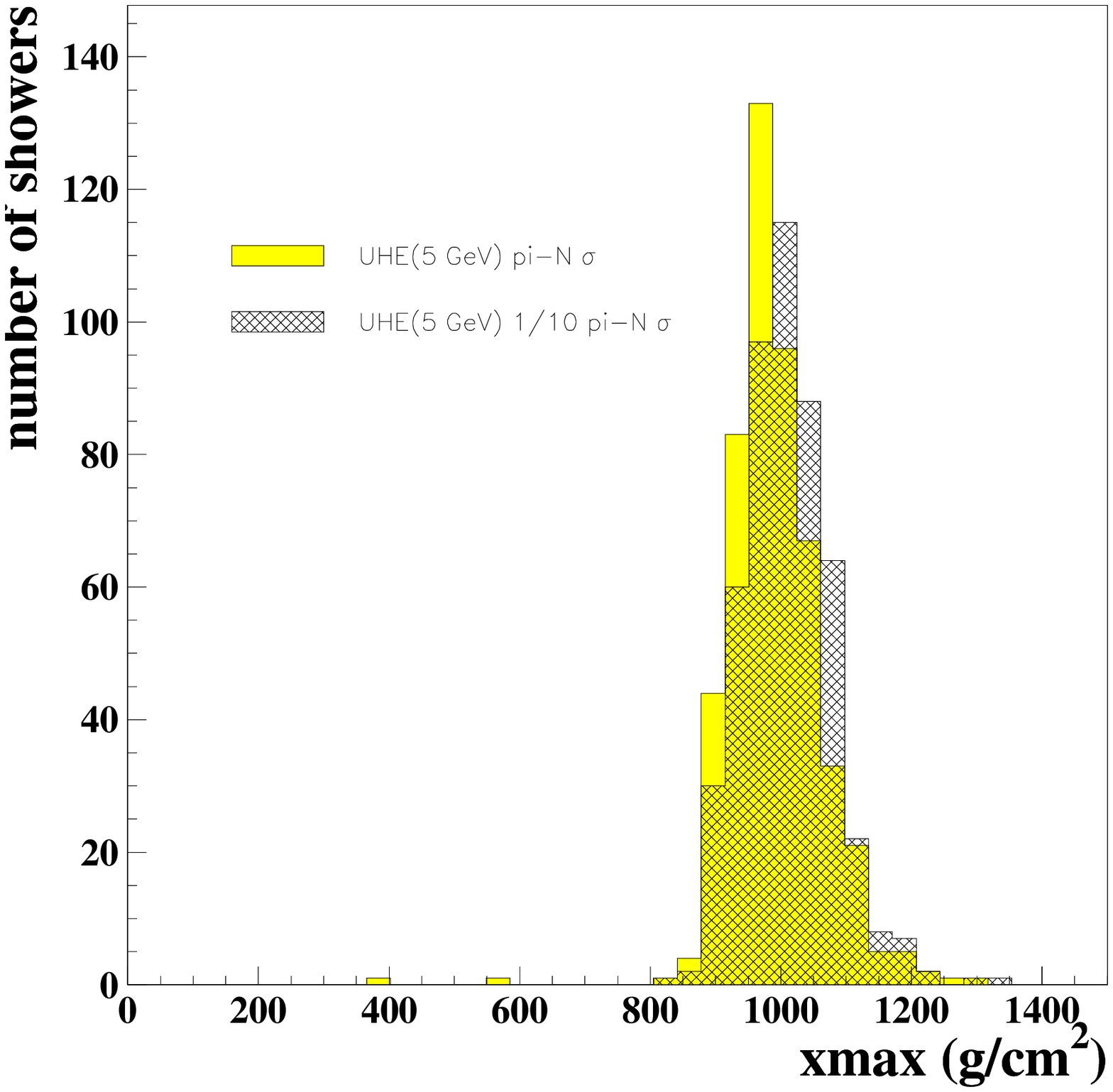}
\epsfxsize=225pt \epsfbox{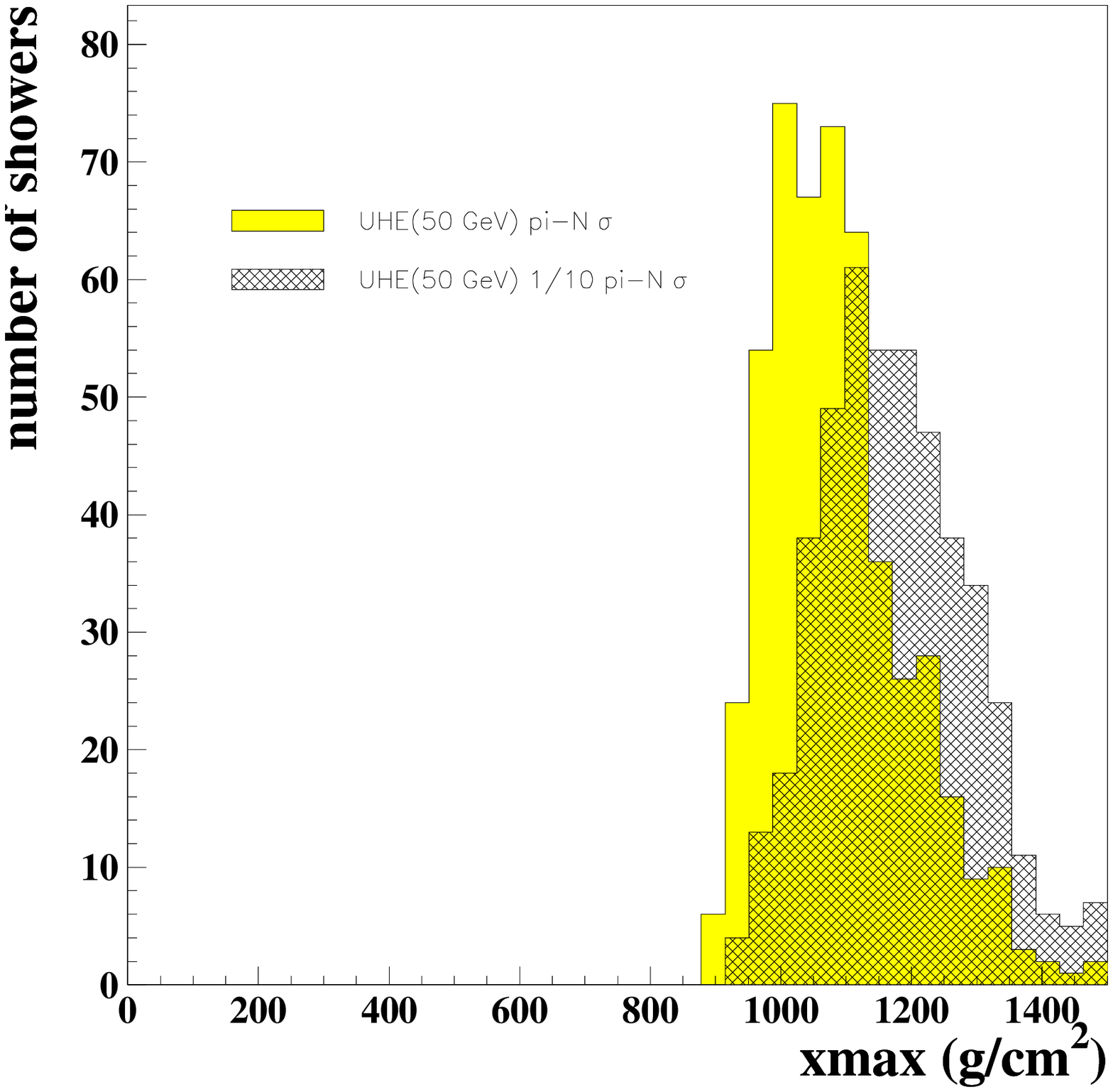}
\Black{
\caption[fig4]{\label{fig:cross} $X_{MAX}$ distributions for primaries
with different cross sections. On the left, a 5 GeV uhecron with
$\sigma$ equal to the pi-N cross section and another with 1/10 of this
value are compared.  On the right, the same for a 50 GeV uhecron.}
}
\end{figure}

\hspace*{1em}
We also investigated the dependence of the longitudinal profile on the
cross section of the uhecron, as discussed in Section \ref{subsec:uhe}.
In addition to using the same pion--nucleon cross section,
we simulated showers where the cross section for the
uhecron--nucleon interaction was 1/10 of the pion--nucleon
interaction. Seven samples were produced, for masses of 2.5, 5, 30, 35, 40,
45 and 50 GeV.  The results of the cross section comparison
are shown in Fig.\ \ref{fig:cross}. As expected, the average $X_{MAX}$
is deeper for smaller cross section. Also the $X_{MAX}$ distribution
is broader.

Analyzing these distributions to determine the maximum uhecron mass
compatible with the observed shower, as in Section \ref{compatibility}
above, shows that if the uhecron-nucleon cross section is lower than a
standard hadronic cross section, the uhecron mass limit is even
stronger.  We find that for a cross section 1/10 the pion--nucleon
cross section, the uhecron mass must be lower than 35 to 40 GeV\footnote{For
completeness, we also produced a sample for a 50 GeV uhecron, assuming $\sigma_{UN}
= 2 \sigma_{\pi N}$, in order to assess the sensitivity to an increase in
cross section.  The change in the $X_{MAX}$ 
distribution is much smaller than in going from $\sigma_{UN}= \sigma_{\pi
N}$ to $\sigma_{UN}= 1/10 \sigma_{\pi N}$; it would have the effect of 
slightly weakening the limit on uhecron mass.  Since we do not consider the
larger cross section as physically well-motivated, we do not pursue this
possibility further.}.

\subsection{Other aspects of the shower development}
\begin{figure}[!t]
\centering 
\leavevmode \epsfxsize=225pt \epsfbox{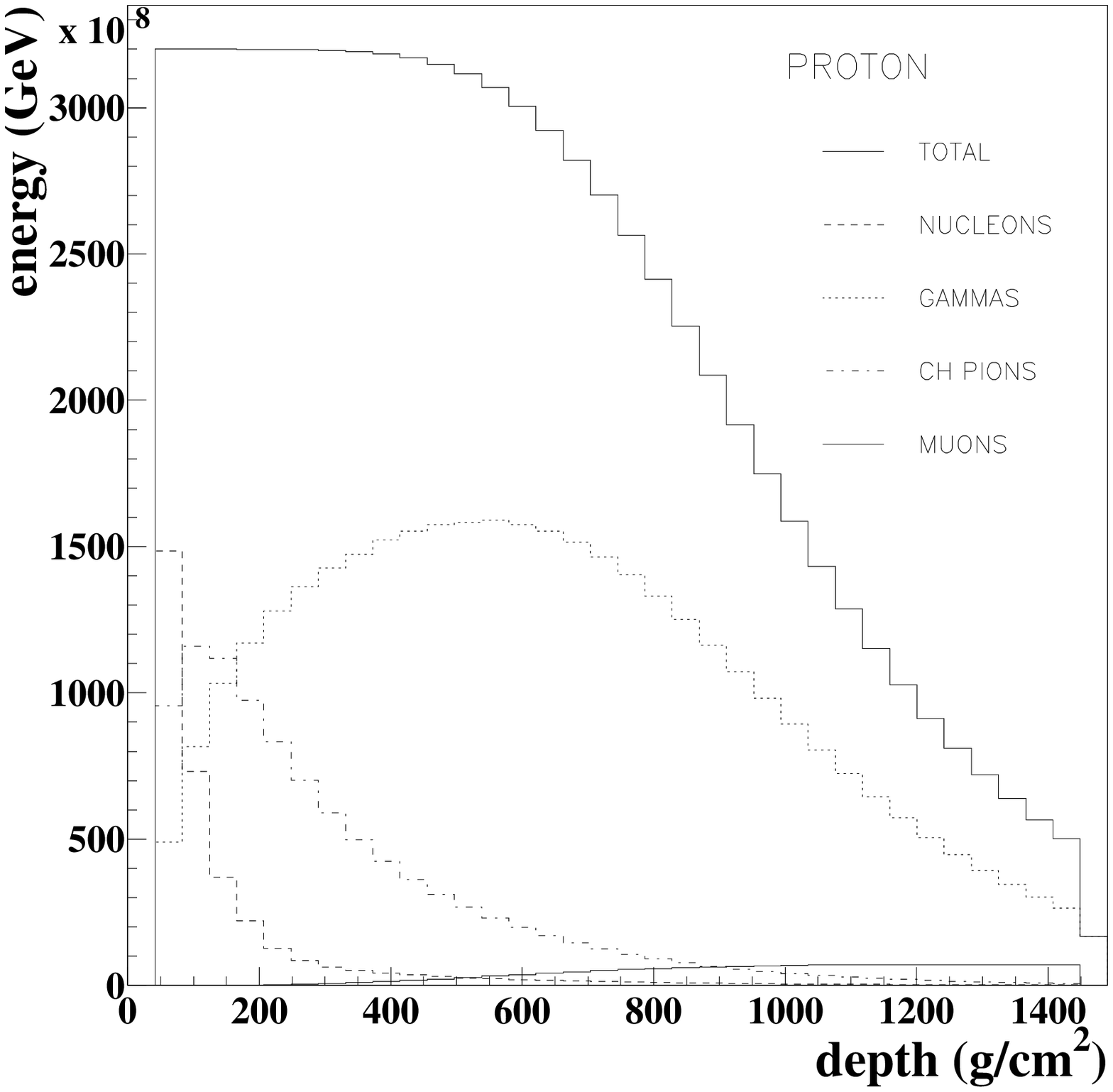}
\epsfxsize=225pt \epsfbox{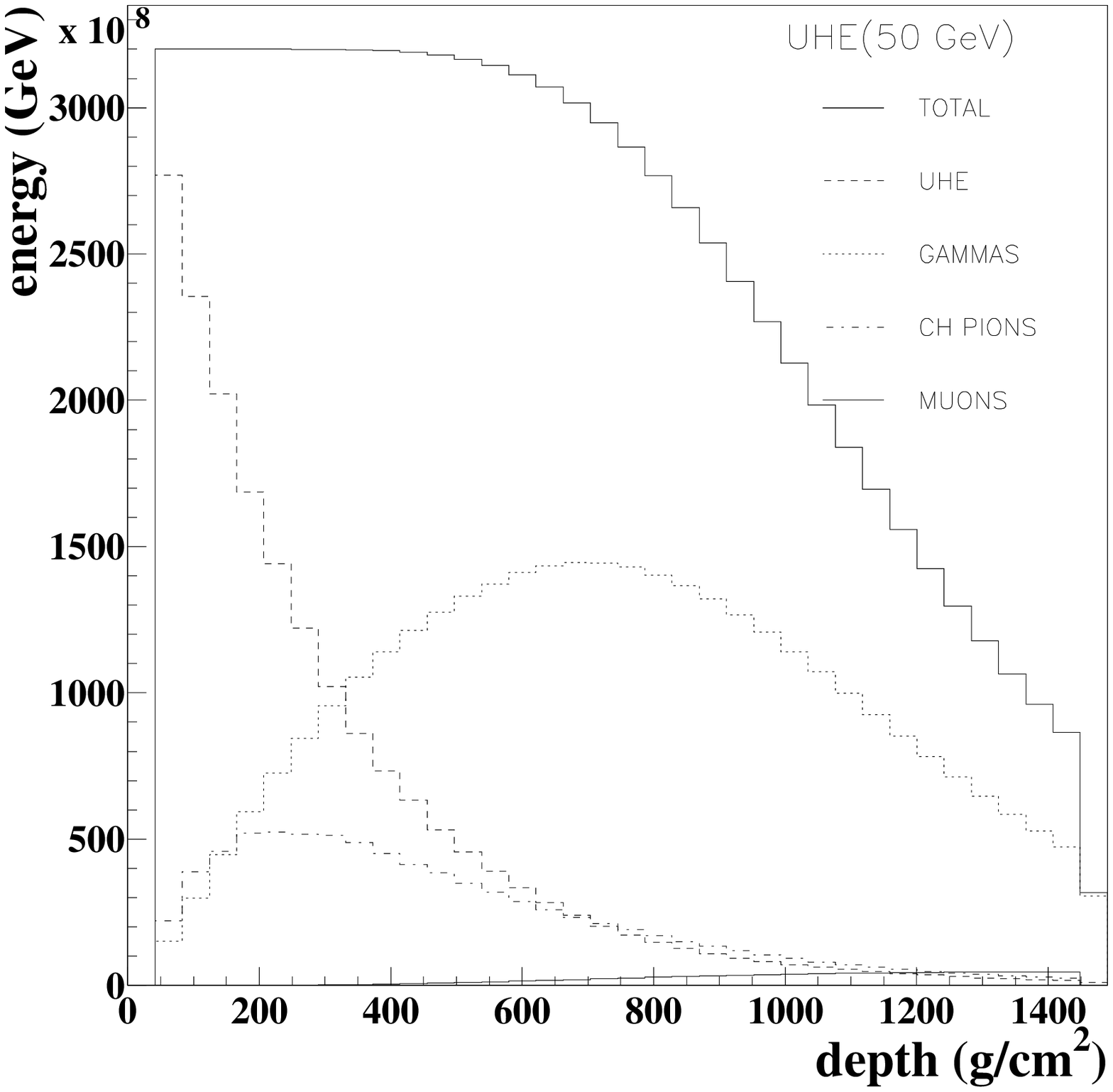}
\caption[fig5]{\label{fig:energy} Distribution of energy among
different constituents of the shower. Shown for the proton
is the energy carried by all particles, by nucleons, gammas, charged pions and
by muons. The same is shown for a 50 GeV
uhecron, with the exception that the energy carried by the
uhecron is shown rather than the energy carried by nucleons.}
\end{figure}

\hspace*{1em} We show in Fig.\ \ref{fig:energy} the energy
distribution among the shower particles for a proton and a 50 GeV
uhecron. The first plot shows how the energy is distributed among
nucleons, gammas, charged pions, and muons in a proton shower. The
second plot shows the same information for a shower initiated by a 50 GeV
uhecron, but showing the energy carried by the uhecron rather than the energy
carried by nucleons.  One sees that the uhecron carries a large fraction
of the energy down to a considerable depth, whereas the energy in the proton
shower is quickly distributed to other particles.  We will return below to
physical consequences of these distributions for using extensive air
shower measurements to distinguish uhecron from proton primaries.

Finally, we also simulated showers for primaries with 100 EeV energy.
As there are no longitudinal distributions observed for events of this
energy, we simply show the results in Table \ref{tab:100e}. The
compatibility among uhecron and proton showers for this primary
energy, can be deduced from the parameters in Table \ref{tab:100e}.
\begin{table}
\caption{The mean of the $X_{MAX}$ distribution and its variance
$\sigma$ for various uhecrons of energy 100 EeV.}
\vspace{12pt}
\centering
\label{tab:100e}
\begin{tabular}{ccc}
\hline \hline
\hspace*{36pt} Primary \hspace*{36pt}& \hspace*{36pt} $\langle X_{MAX}
\rangle$(g\,cm$^{-2}$) \hspace*{36pt} & \hspace*{36pt}
$\sigma$(g\,cm$^{-2}$)\hspace*{36pt} \\ \hline 
Proton & 884 & 51\\
uhecron (5 GeV) & 949 & 59\\ 
uhecron (50 GeV) & 1031 & 79\\
\hline\hline
\end{tabular}
\vspace{4mm}
\end{table}

\subsection{Ground level particle content}
\hspace*{1em}
In order to find a signature for the uhecron in an extensive air
shower array detector, we studied the ratio between the muon (mu)
and the electromagnetic (em) density at ground level. 
To understand fully if this ratio can be used to distinguish a uhecron
from other primaries, a complete analysis incorporating a model 
of the detector as well as thresholds and efficiencies would be needed.
However our estimation can give an idea of the feasibility of this
kind of signature.

\begin{figure}[!t]
\centering 
\leavevmode \epsfxsize=225pt \epsfbox{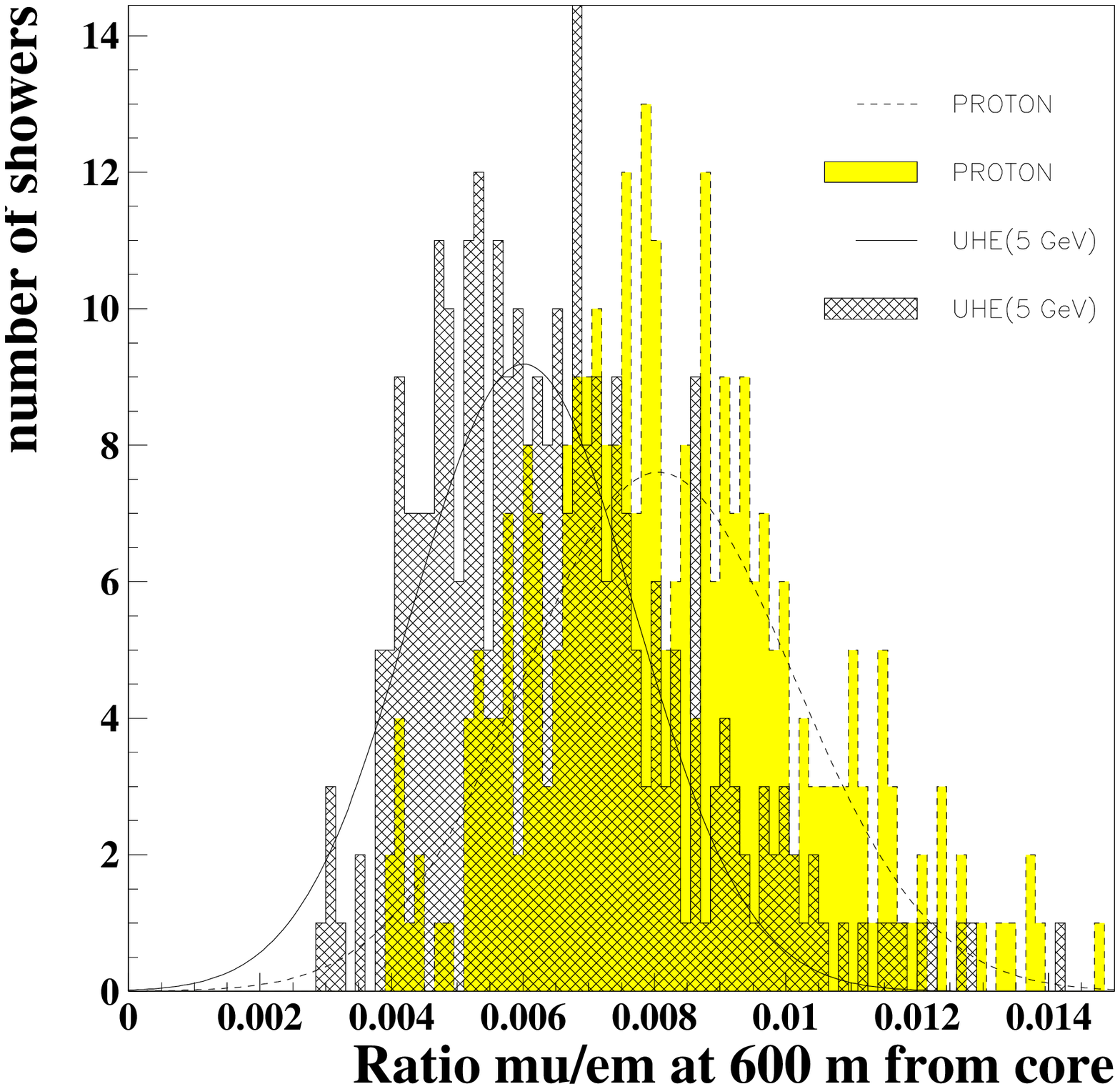}
\epsfxsize=225pt \epsfbox{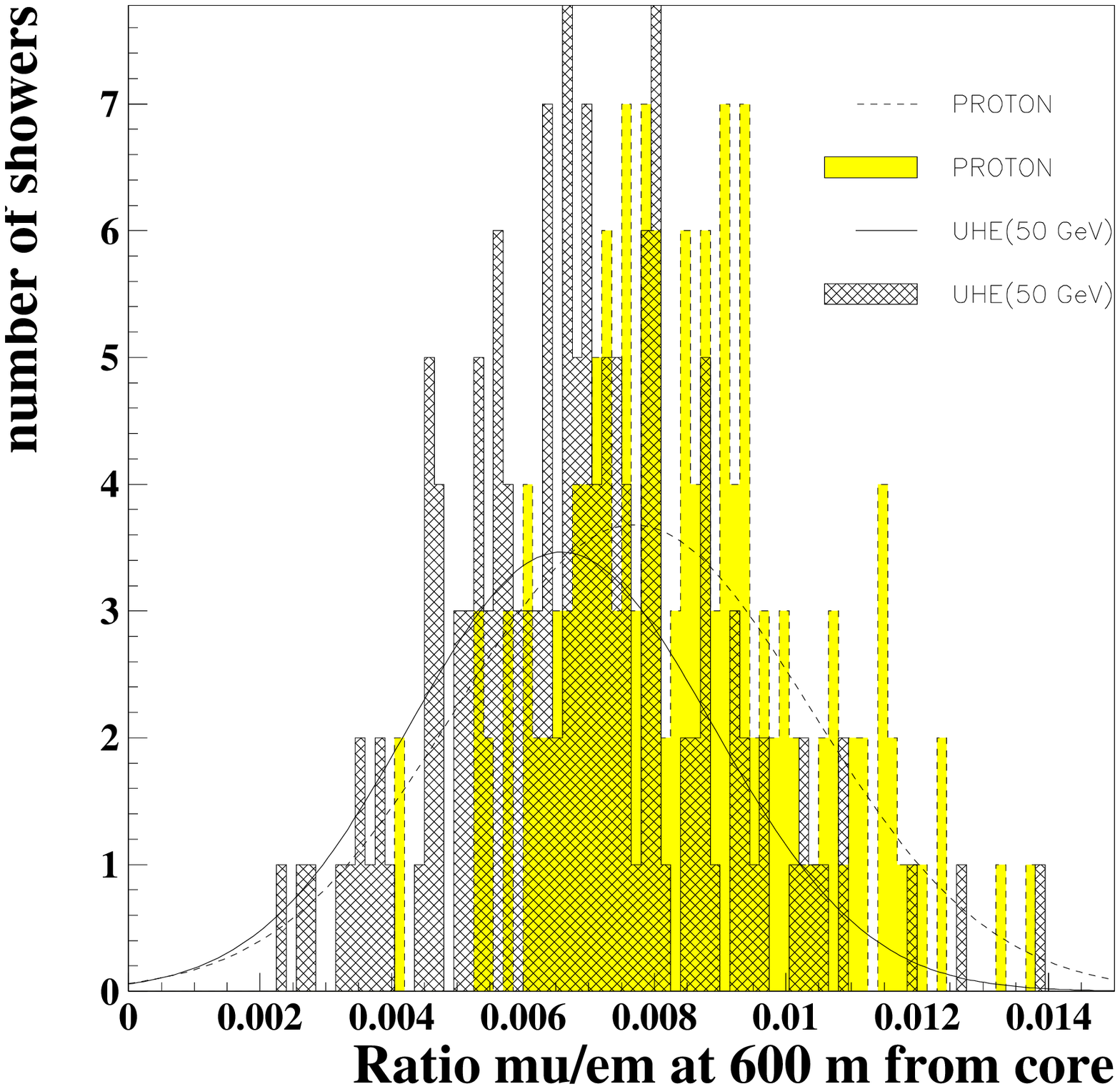}
\caption[fig6]{\label{fig:mu} The muon to electromagnetic density
ratio at ground level.  The left figure is the result of 300 showers
each for a 5 GeV uhecron and a proton, while the right figure is 150
showers each for a 50 GeV uhecron and a proton.}

\end{figure}

The zenith angle for this analysis is chosen to be $30^\circ$ and ground level
is taken to be 870 g\,cm$^{-2}$. The distribution of the ratio mu/em at 600 
meters from the core is plotted in Fig.\ \ref{fig:mu}. The figure shows this 
ratio for a 5 GeV uhecron and for a proton, and for a 50 GeV uhecron and a 
proton.

{}From Fig.\ \ref{fig:mu} one can see that the the mu/em ratio has a
large dispersion. But the first plot shows there seems to be a qualitative
difference between even a 5 GeV uhecron and a proton: 
the proton mu/em ratio is larger than the 5 GeV uhecron mu/em ratio.
This difference can be understood by analyzing the muonic
and electromagnetic longitudinal development, keeping in mind
that the proton interaction with the atmosphere is very different from the
uhecron interaction.  At the top of the atmosphere, the particles produced
by a proton shower are more energetic than the ones produced in a uhecron 
shower, since the light, strongly-interacting quanta carry a larger fraction
of the initial energy in the former case. This can be seen from the plots
of Fig.\ \ref{fig:energy}.

The analysis of the muonic and electromagnetic longitudinal development
shows that the difference in the number of muons between a proton and
a uhecron shower is larger than the difference between the number of 
electromagnetic particles. This is due to the fact that the proton shower
produces much more energetic pions in the top of the atmosphere. The
charged pions therefore have enough time to interact, multiplying the number
of pions, before decaying into muons. In the uhecron shower the charged pions
at the 
top of the atmosphere are less energetic and will most likely decay directly 
into muons before interacting. This is not true for the electromagnetic
component of the shower, which behaves in the same way in a proton and
a uhecron shower, since a $\pi^{0}$ always decays rather than interacts.

The second plot in Fig.\ \ref{fig:mu} is for a 50 GeV uhecron. At first
sight it is surprising that the uhecron mu/em ratio is closer to the proton
ratio than the 5 GeV uhecron. This seems to be counter to the above explanation
since the pions in the top of the atmosphere in the 50 GeV shower should be 
even less energetic than the ones in a 5 GeV shower. The effect, however,
is due to an artifact of our choices of the ground level (made in order to
be consistent with experimental sites) and the zenith angle for this analysis.

With the ground level depth at 870 g\,cm$^{-2}$ and a $30^\circ$ zenith
angle, the atmospheric depth is about 1005 g\,cm$^{-2}$. This is enough
for both a proton and a uhecron shower to have passed their maxima, while
it is not enough for a 50 GeV uhecron shower (see Fig.\ \ref{fig:prof}). 
Looking more closely at the muonic and electromagnetic longitudinal development,
reveals that the muonic component of a 50 GeV uhecron shower has not yet reached its 
maximum when arriving at ground level, whereas the electromagnetic component
is almost at it's maximum at 1005 g\,cm$^{-2}$. This is confirmed when
one chooses a deeper ground level, where the
50 GeV mu/em ratio is smaller than the 5 GeV ratio.

An important point, evident from the difference between the fits to
300 and 150 proton showers in Fig.\ \ref{fig:mu}, is that very large
statistics are essential to use mu/em as a tool for discriminating a
uhecron from a proton.  This, combined with the small shift in going
from proton to uhecron, means that sensitivity to details of the
shower modeling must be understood before the calculated difference
can be trusted.

If both longitudinal shower development and the mu/em
ratio is available for a given event, correlated features may provide
greater resolving power. As shown in a comparison between iron and
proton showers \cite{auger}, the mu/em ratio and the XMAX
measurement are correlated and dependent on shower development parameters.
Thus with large statistics, a multidimensional fit can be more
sensitive than either alone.

\section{Discussion and Conclusion}
\label{sec:conc}
\hspace*{1em} We conclude from our simulations that a uhecron (a
massive, long-lived and strongly interacting particle) that penetrates
Earth's atmosphere can produce an air shower with characteristics
qualitatively similar to those of a proton shower.  We find an
approximate upper limit of 50 GeV on the mass of such a particle in
order to have a shower compatible with the longitudinal development
observed in the highest energy cosmic ray event \cite{flys}.  If the
cross section of the uhecron is lower than a ``normal" hadronic cross
section, this limit is even more stringent.  For instance it drops to
40 GeV for a cross section 1/10 of the pion-nucleon cross section.

We note that as the mass of the uhecron increases, the shower develops
more slowly (larger average $X_{MAX}$) and the fluctuations in
$X_{MAX}$ increase.  With a large number of events whose longitudinal shower
development is measured, these differences could be sufficient to distinguish
uhecron from proton primaries.  We estimate that about 30 events may be enough
to discriminate between uhecron and proton, if the uhecron is more massive
than 10 GeV. At lower uhecron mass the discrimination is difficult because
the longitudinal development profiles become too similar.  In this case the
ratio of muon over electromagnetic density at large distances from the core
could be helpful with very high statistics.  We conjecture that having
combined information on a single event, longitudinal development as well as
lateral shower properties and the mu/em ratio, may give improved
resolving power.  Such combined information would be available if the
event were observed simultaneously with both fluorescence and ground
level detectors, as in the Pierre Auger project \cite{auger}.

It is encouraging that approved and proposed experimental efforts
\cite{auger,hires,teles} are likely to produce a substantial increase
in the statistics available on ultrahigh energy cosmic rays.  We have
shown that the detailed shower properties can allow candidate
new-particle explanations for the highest energy events to be ruled
out or conceivably confirmed.  Many possible new-particle explanations
are already excluded.  In particular, the highest energy cosmic ray
event is very unlikely to have been produced by a new superheavy ($M
\ga 100$ GeV) hadron containing a heavy colored particle or monopole.

\vspace{24pt} We would like to thank Clem Pryke, Jim Cronin, Gustavo
Burdman, Sergio Sciutto and Albert Stebbins for useful discussions.
IA was supported in part by NSF Grant AST 94-20759 and the DOE through
grant DE-FG02-90ER40606. The work of GRF was supported by
NSF-PHY-94-23002.  The work of EWK was supported at Fermilab by the
Department of Energy and by NASA under number NAG5-7092.

\end{document}